\documentstyle[prc,aps,epsfig]{revtex}

\tightenlines

\begin{document}
%\draft
%\def\pmb#1{\setbox0=\hbox{#1}%
%     \kern-.025em\copy0\kern-\wd0
%      \kern.05\copy0\kern-\wd0
%       \kern-.025em\raise.0433em\box0}
%\def\btau{\pmb{$\tau$}}
%\def\bsigma{\pmb{$\sigma$}}
%\def\bdelta{\pmb{$\delta$}}
%\def\bdiamond{\pmb{$\diamond$}}
%\def\bbdiamond{\pmb\bdiamond}
%\def\hat{\widehat}
%\def\omeg{\bmatdieci\char '41}

\date{\today}
%\begin{frontmatter}
\title
{Collective modes of asymmetric nuclear matter in Quantum HadroDynamics}
\author{
V.Greco${}^{{1}}$, M.Colonna${}^{{1}}$, 
M.Di Toro${}^{{1}}$, F.Matera${}^{{2}}$ }

\address
{$^{1}$Laboratorio Nazionale del Sud, Via S. Sofia 44,
I-95123 Catania, Italy\\
and Dipartimento di Fisica, Universit\`a degli Studi di Catania\\}

\address
{$^{2}$Dipartimento di Fisica, Universit\`a degli Studi di Firenze\\
and Istituto Nazionale di Fisica Nucleare, Sezione di Firenze,\\
Via G. Sansone 1, I-50019, Sesto F.no (Firenze), Italy} 

\maketitle

\begin{abstract}
We discuss a fully relativistic Landau Fermi liquid theory based on the 
Quantum Hadro-Dynamics ($QHD$) effective field picture of Nuclear Matter
 ({\it NM}). From the linearized kinetic equations we get the dispersion
relations of the propagating collective modes. We focus our attention on the 
dynamical effects of the interplay between scalar and vector channel 
contributions. A beautiful ``mirror'' structure in the form of the 
dynamical response in 
the isoscalar/isovector degree of freedom is revealed, with a complete 
parallelism in the role respectively played by the compressibility and the
symmetry energy. All that strongly supports the introduction of an explicit 
coupling to the scalar-isovector channel of the nucleon-nucleon interaction.

In particular we study the influence of this coupling (to a $\delta$-meson-like
effective field) on the collective response of asymmetric nuclear 
matter ($ANM$).
Interesting contributions are found on the propagation of isovector-like
modes at normal density and on an expected smooth transition to isoscalar-like
oscillations at high baryon density.

Important ``chemical'' effects on the neutron-proton structure of the mode are
shown. For dilute $ANM$ we have the isospin distillation mechanism of the 
unstable isoscalar-like oscillations, while at high baryon density we predict
an almost pure neutron wave structure of the propagating sounds.
\end{abstract}


\vskip 1.0cm

\section{Introduction}
The $QHD$ effective field model represents a very succesfull attempt to 
describe, in a fully consistent relativistic picture, equilibrium and 
dynamical properties of nuclear systems at the hadronic 
level \cite{wal74,Se86,Se97}.
Very nice results have been obtained for the nuclear structure of finite 
nuclei \cite{snr94,ri96,tywo99}, for the $NM$ Equation of State and liquid-gas 
phase transitions \cite{mu95} and for the dynamics of nuclear collisions
\cite{bla93,koli96}. Relativistic Random-Phase-Approximation ($RRPA$) 
theories have been developed to study the nuclear collective response 
\cite{de91,ma94,vre97,magi97,magi01,rpa00}.

In this paper we present a relativistic linear response theory with the aim
of a trasparent connection between the collective dynamics and the 
coupling to various channels of the nucleon-nucleon interaction. In 
particular we will focus our attention on the dynamical response of 
asymmetric nuclear matter
since one of the main points of our discussion is the relevance of the coupling
to a scalar isovector channel, the virtual $\delta[a_0(980)]$ meson, not
considered in the usual dynamical studies. Another point of interest is the 
dynamical treatment of the Fock terms, neglected in the usual 
Relativistic Mean Field ($RMF$) scheme of the papers cited before.

A relativistic extension of the Landau linear response theory of Fermi Liquids
has been considered before just starting from the relativistic form of the
Landau parameters \cite{mat81,song,cai01}. We will show that the full 
dispersion relations obtained from the relativistic kinetic equations 
present some interesting corrections that cannot be neglected.

The main physics results are:
\begin{itemize}
\item The important effect of a $\delta$-meson coupling on the 
isovector collective mode at saturation baryon density. This is of 
interest for the relativistic study of the Giant Dipole Resonance 
in heavy finite nuclei. It is important
 to note that the inclusion of Fock terms is acting in the same direction.
\item The presence of noticeable ``chemical effects'' in the propagating 
collective oscillations, i.e. the charge symmetry of the ``waves'' is 
quite different from the asymmetry of the initial equilibrium matter. 
The effect is opposite for the
 unstable modes present at low densities, more proton rich and 
leading to the isospin distillation effect, and for the stable 
propagating sounds at high 
baryon density which appear mostly like ``pure neutron waves''.
\item A stimulating ``mirror'' structure of the isoscalar and isovector 
linear response, with the restoring forces given by the potential part 
respectively of the compressibility and the symmetry energy.
The interplay between the scalar and vector meson effective fields in 
the dynamics is very similar for the two degrees of freedom, as already 
observed for static properties \cite{liu02}. The conclusion is that a 
consistent relativistic effective field model has to include on the same 
footing isoscalar and isovector meson fields, $both$ scalar and vector.
\end{itemize} 
Our results around normal density can be used as general guidelines in 
predicting the behaviour of volume collective modes in finite
$\beta$--unstable nuclei.  
Similar study for asymmetric $NM$ have been performed in Ref.
\cite{lar98} using Skyrme--like
interactions. Apart the difference in the used interactions, in particular 
for the symmetry terms, we will see similar results and interesting 
new relativistic effects.

In Sect.II we derive the kinetic equations in the general case of non-linear
self-interacting terms, including the Fock corrections. We discuss the 
inclusion of the $\delta$-meson channel, also on the model parameters. In 
Sect.III we present the relativistic linear response equations. In Sect.IV 
we have a general discussion on the formal structure of the dispersion 
relations, the role played by the scalar/vector mesons and the 
comparison to non-relativistic cases.
Results for isovector(-$like$) collective modes are presented in Sect.V, in 
particular for the asymmetry and baryon density effects. The isoscalar(-$like$)
response is analysed in Sect. VI. Conclusions and 
outlooks can be found in Sect.VII.

\section{Kinetic Equations from a QHD effective theory}

We start from the {\it $QHD$} effective field picture of the hadronic 
phase of nuclear matter \cite{wal74,Se86,Se97}. In order to include the 
main dynamical 
degrees of freedom of the system we will consider the nucleons
coupled to the isoscalar scalar $\sigma$ and vector $\omega$ mesons
and to the isovector scalar $\delta$ and vector $\rho$ mesons.

The Lagrangian density for this model, including non--linear isoscalar/scalar 
$\sigma$-terms \cite{bo89}, is given by:
\begin{eqnarray}\label{eq.1}
{\cal L} = {\bar {\psi}}[\gamma_\mu(i{\partial^\mu}-{g_\omega}{\cal V}^\mu
 - g_{\rho}{\bf {\cal B}}^\mu \cdot {\pmb{$\tau$}} ) -
(M-{g_\sigma}\phi-g_{\delta}{\pmb{$\tau$}}\cdot{\pmb{$\delta$}})]\psi + 
\nonumber\\
  {1\over2}({\partial_\mu}\phi{\partial^\mu}\phi
- m_s^2 \phi^2) - {a \over 3} \phi^3 - {b \over 4} \phi^4
- {1 \over 4} W_{\mu\nu}
W^{\mu\nu} + {1 \over 2} m_v^2 {\cal V}_\nu {{\cal V}^\nu}+ \nonumber\\
\frac{1}{2}(\partial_{\mu}{\pmb{$\delta$}}\cdot\partial^{\mu}{\pmb{$\delta$}}
-m_{\delta}^2{\pmb{$\delta$}}^2)
- {1 \over 4} {\bf G}_{\mu\nu}\cdot
{\bf G}^{\mu\nu} + {1 \over 2} m_{\rho}^2 {\bf {\cal B}}_\nu\cdot
{{\bf {\cal B}}^\nu}         
\end{eqnarray}
where
$W^{\mu\nu}(x)={\partial^\mu}{{\cal V}^\nu}(x)-
{\partial^\nu}{{\cal V}^\mu}(x)~$ and
${\bf G}^{\mu\nu}(x)={\partial^\mu}{{\bf {\cal B}}^\nu}(x)-
{\partial^\nu}{{\bf {\cal B}}^\mu}(x)~.$

Here $\psi(x)$ is the nucleon
fermionic field, $\phi(x)$ and ${{\cal V}^\nu}(x)$ represent neutral scalar
 and
vector boson fields, respectively. ${\pmb{$\delta$}}(x)$ and 
${{\bf {\cal B}}^\nu}(x)$
are the charged scalar and vector
fields and ${\pmb{$\tau$}}$ denotes the isospin matrices .

From the Lagrangian Eq.(\ref{eq.1}) with the Euler procedure a set of 
coupled equations of motion for the meson and nucleon fields can be 
derived. The basic 
approximation in nuclear matter applications consists in neglecting all the
terms containing derivatives of the meson fields, with respect to the mass
contributions. Then the meson fields are simply connected to the operators
of the nucleon scalar and current densities by the following equations: 
\begin{equation}\label{Eq.4a}
{\widehat{\Phi}/f_\sigma} + A{\widehat{\Phi}^2}
+ B{\widehat{\Phi}^3}~=\bar\psi(x)\psi(x)~\equiv \widehat{\rho_S}
\end{equation}
\begin{eqnarray}\label{Eq.4b}
{\widehat{\cal V}}^\mu(x)&=f_\omega\bar\psi(x){\gamma^\mu}\psi(x)~
\equiv f_\omega \widehat j_\mu~,\nonumber\\
{\widehat  {\bf B}}^\mu(x)&=f_{\rho}\bar\psi(x){\gamma^\mu} {\pmb{$\tau$}}
\psi(x)~,\nonumber\\
{\widehat{\pmb{$\delta$}}}(x)&=f_{\delta}\bar\psi(x){\pmb{$\tau$}}\psi(x)
\end{eqnarray}
where $\widehat \Phi=g_\sigma\phi$, $f_\sigma = (g_\sigma/m_\sigma)^2$, 
$A = a/g_\sigma^3$, $B = b/g_\sigma^4$, $f_\omega = (g_\omega/m_\omega)^2$, 
$f_{\rho} = (g_{\rho}/2m_{\rho})^2$, $f_{\delta} = (g_{\delta}/m_{\delta})^2$. 

\par
For the nucleon fields we get a Dirac-like equation. Indeed after 
substituting Eqs.(\ref{Eq.4a},\ref{Eq.4b}) for the meson field operators, we 
obtain an equation which contains only nucleon field operators. All the 
equations can be consistently solved in a Mean Field Approximation 
($RMF$), where most
applications have been performed, in particular in the Hartree scheme
\cite{Se97,ri96}.

The inclusion of Fock terms is conceptually important \cite{hor83,bou87}
since it automatically leads to contributions to various channels, also
in absence of explicit coupling terms. We will discuss this point later.
A thorough study of the Fock contributions in a $QHD$ approach with non-linear
self-interacting terms has been recently performed \cite{prc_nuovo}, in
particular for asymmetric matter \cite{gre01}.
 
The present approximation implies that retardation and finite range
effects in the exchange of mesons between nucleons are
neglected. 
Nevertheless, thanks to the small Compton wave--lengths
of the mesons $\sigma$, $\omega$,$\rho$ and $\delta$, the assumptions
expressed by Eqs. (\ref{Eq.4a},\ref{Eq.4b}) are quite reasonable.
For light mesons such as pions this approximation is not justified.
 However in this case a perturbative expansion
in the pion-nucleon coupling constant seems to be reasonable \cite{hor83}.
Moreover it has been shown that
the inclusion of pions does not change qualitatively the description of
nuclear matter around normal conditions \cite{hor83}.

We remark that the kinetic approach discussed here is fully consistent
with the previous approximation. We are concerned with a semiclassical 
description of nuclear dynamics, so that the nuclear medium is supposed
to be in states for which the nucleon scalar and current densities
are smooth functions of the space-time coordinates.

Within a mean field picture of the $QHD$ model
we focus our analysis on a description of 
the many-body nuclear system in terms of one--body dynamics.
This is enough for the scope of the paper. Correlation effects can be 
effectively included at the level of coupling constants, as noted in the 
discussion of the results.

We will perform the many-body calculations in the quantum phase-space
introducing the Wigner transform of the one-body density matrix for the
fermion field \cite{degr80,hak78}.

The one--particle Wigner function is defined as:
$$[{\widehat F}(x,p)]_{\alpha\beta}=
{1\over(2\pi)^4}\int d^4Re^{-ip\cdot R}
\langle :\bar{\psi}_\beta(x+{R\over2})\psi_\alpha(x-{R\over2}):\rangle~,$$
where $\alpha$ and $\beta$ are double indices for spin and isospin. 
The brackets denote statistical averaging and the colons denote 
normal ordering. 
The Wigner function is a matrix in spin and isospin spaces; in the case of
asymmetric NM it is useful to decompose it into neutron and proton components.
Following the treatment of the Fock terms in non-linear $QHD$ introduced 
in Ref. \cite{prc_nuovo,gre01}, we obtain for the Wigner function the 
following kinetic equation: 
\begin{eqnarray}\label{trans}
&&{i\over2}{\partial_\mu}{\gamma^\mu}{\hat F}^{(i)}(x,p)+\gamma^\mu
p^*_{\mu i}{\hat F}^{(i)}(x,p) - M^*_i{\hat F}^{(i)}(x,p)+\nonumber\\
&&{i\over2}\Delta
\left[\tilde f_\omega j_\mu(x)\gamma^\mu \pm \tilde f_\rho j_{3\mu}(x)
\gamma^{\mu}
- \tilde f_\sigma \rho_S(x) \mp \tilde f_\delta \rho_{S3}(x)\right]
{\hat F}^{(i)}(x,p)=0,~~~~~~~i=n,p
\end{eqnarray}
where $\Delta={\partial_x}\cdot{\partial_p}$, with $\partial_x$ acting only 
on the first term of the products.
Here $\rho_{S3}=\rho_{Sp}-\rho_{Sn}$ and $j_{3\mu}(x)=j^p_{\mu}(x)
-j^n_{\mu}(x)$ 
are the isovector scalar density and the isovector baryon current, 
respectively. 
We have defined the kinetic momentum and effective masses, as:
\begin{eqnarray}\label{psms}
&&p^*_{\mu i}=p_\mu-{\tilde f_\omega}j_\mu(x)\pm \tilde f_\rho{j}_{3\mu}(x)\nonumber\\
&&M^*_i=M- {\tilde f_\sigma}\rho_S(x) \pm \tilde f_\delta \rho_{S3}(x)~, 
\end{eqnarray}
with the effective coupling functions given by:
\begin{eqnarray}\label{cincoup}
&&{\tilde f}_\sigma={{\Phi} \over {\rho_S}}-{1\over 8}{{d\Phi(x)} \over {d\rho_S(x)}}
- {1\over {2\rho_S}}Tr {\widehat F}^{2}(x)
{{d^{2}\Phi(x)} \over {d\rho_S^{2}(x)}}
+{1\over 2}f_\omega+{3\over 2}f_{\rho}-{3\over 8}f_{\delta}~,\nonumber\\
&&{\tilde f}_\omega={1\over 8}{{d\Phi(x)} \over {d\rho_S(x)}}
+{5\over 4}f_\omega+{3\over 4}f_{\rho}+{3\over 8}f_{\delta}~,\nonumber\\
&&\tilde f_\delta=-{1\over 8}{{d\Phi(x)} \over {d\rho_S(x)}}
+{1\over 2}f_\omega-{1\over 2}f_{\rho}+{9\over 8}f_{\delta}~,\nonumber\\
&&\tilde f_\rho={1\over 8}{{d\Phi(x)} \over {d\rho_S(x)}}
+{1\over 4}f_\omega+{3\over 4}f_{\rho}-{1\over 8}f_{\delta}~,\nonumber\\
\end{eqnarray}
where $8~Tr {\widehat F}^{2}(x)=\rho_S^2+j_\mu j^\mu+\rho_{S3}^2+
j_{3\mu} j^{3\mu}$.
We remind that we are dealing with a 
transport equation so the 
currents and densities, in general, are varying functions 
of the space--time, at variance with the case of nuclear matter at equilibrium. 

The expression of Eq.(\ref{psms}) for the effective mass, 
embodies an isospin contribution from Fock terms also without 
a direct inclusion of 
the $\delta$ meson in the Lagrangian. 
The usual $RMF$ approximation (Hartree level) is covered by the 
Hartree-Fock results, one has has only to change the coupling functions 
$\tilde f_i (i=\sigma,\omega, \rho, \delta)$, Eqs.(\ref{cincoup}), with 
the coupling constants $f_i$.

\subsection*{Equilibrium properties: the nuclear Equation of State}

In the following we will study the collective modes. In order to analyze
the results it is essential to relate them to the equation of state ($EOS$),
that we will 
briefly discuss in the following.
In particular for the collective modes in asymmetric nuclear matter 
it is important 
the behaviour of the symmetry energy $E_{sym}$.

The energy density and pressure for symmetric and asymmetric 
nuclear matter and the $n,p$ effective masses 
can be self-consistently calculated  
 just in terms of the four boson coupling constants,
$f_i \equiv (\frac{g_i^2}{m_i^2})$, $i = \sigma, \omega, \rho, \delta$,
and the two parameters of the $\sigma$ self-interacting terms, 
$A \equiv \frac{a}{g_\sigma^3}$ and $B \equiv \frac{b}{g_\sigma^4}$,
see ref.\cite{prc_nuovo,gre01}. 

The isoscalar meson parameters are fixed from symmetric nuclear matter
properties at $T=0$: saturation density $\rho_0=0.16fm^{-3}$,
 binding energy $E/A = -16MeV$, nucleon effective mass $M^\star = 0.7 M_N$
 ($M_N=939MeV$) and incompressibility $K_V = 240 MeV$ at $\rho_0$. The fitted
$f_\sigma, f_\omega, A, B$ parameters are reported in Table I. They have
quite standard values for these minimal non-linear $RMF$ models. $Set~I$ and 
$Set~II$ correspond to the best parameters within a non-linear Hartree 
calculation, respectively with the $\rho-(Set~I, NLH+\rho)$ and with the
$\rho+\delta$ $(Set~II,NLH+(\rho+\delta))$ couplings in the isovector channel
(see the discussion in ref.\cite{liu02}). $NLHF$ stands for the 
non-linear Hartree-Fock scheme described before.

\vspace{1cm}
\noindent

{\bf Table I.} Parameter sets.

\par
\vspace{0.8cm}
\noindent
 
\begin{center}
\begin{tabular}{ c c c c c } \hline
$parameter$   & $Set~I$         &~ $Set~II$    &~$NLHF$   &~ $NL3$  \\ \hline
$f_\sigma~(fm^2)$  &11.27    &$same$           &~9.15     &~ 15.73  \\ \hline
$f_\omega~(fm^2)$  &6.48    &$same$            &~3.22     &~ 10.53  \\ \hline
$f_\rho~(fm^2)$    &1.0     &~2.8             &~1.9      &~ 1.34   \\ \hline
$f_\delta~(fm^2)$  &0.00     &~2.0            &~1.4      &~ 0.00   \\ \hline
$A~(fm^{-1})$     &0.022    &$same$            &~0.098    &~ -0.01  \\ \hline
$B$               &-0.0039  &$same$            &~-0.021   &~ -0.003 \\ \hline
\end{tabular}
\end{center}

\vskip 1cm

In the table we report also the $NL3$ parametrization, widely used in
nuclear structure calculations \cite{lari97}. We remind that the
$NL3$-saturation properties for symmetric matter are chosen as 
$\rho_0 = 0.148fm^{-3}$, $M^* = 0.6 M_N$, 
$K_V = 271.8 MeV$. The symmetry parameter
is $a_4 = 37.4 MeV$.

The symmetry energy in $ANM$ is defined from the expansion
of the energy per nucleon $E(\rho_B,\alpha)$ in terms of the asymmetry
parameter $\alpha$ defined as 
$$\alpha\equiv -\frac{\rho_{B3}}{\rho_B}=\frac{\rho_{Bn}-\rho_{Bp}}
{\rho_B}=\frac{N-Z}{A}.$$
We have
\begin{equation}\label{eq.7}
E(\rho_B,\alpha)~\equiv~\frac{\epsilon(\rho_B,\alpha)}{\rho_B}~
=~E(\rho_B) + E_{sym}(\rho_B) \alpha^2 \nonumber \\
+ O(\alpha^4) +...
\end{equation}
and so in general
\begin{equation}\label{eq.8}
E_{sym}~\equiv~\frac{1}{2} \frac{\partial^2E(\rho_B,\alpha)}
{\partial\alpha^2} \vert_{\alpha=0}~=~\frac{1}{2} \rho_B 
\frac{\partial^2\epsilon}{\partial\rho_{B3}^2}
 \vert_{\rho_{B3}=0}
\end{equation}

In the Hartree case an explicit expression for the symmetry energy can be 
easily derived \cite{ku97,liu02}
\begin{equation}\label{eq.9}
E_{sym}(\rho_B) = \frac{1}{6} \frac{k_F^2}{E^*_F} + \frac{1}{2}
f_\rho \rho_B - \frac{1}{2} f_\delta 
\frac{{M^*}^2 \rho_B}{E_F^{* 2} \left[1+f_\delta A(k_F,M^*)\right]}
\equiv E_{sym}^{kin}+ E_{sym}^{pot}
\end{equation}
where $k_F$ is the nucleon Fermi momentum corresponding to $\rho_B$, 
$E^*_F \equiv \sqrt{(k_F^2+{M^*}^2)}$ and $M^*$ is the effective
nucleon mass in symmetric $NM$, $M^*=M_N - g_\sigma \phi$.

The integral
\begin{equation}\label{eq.10}
A(k_F,M^*)~\equiv~\frac{4}{(2\pi)^3} \int d^3k \frac{k^2}{ 
(k^2+{M^*}^2)^{3/2}}=3 \left(\frac{\rho_S}{M^*} - 
\frac{\rho_B}{E^*_F}\right)
\end{equation}

We remark that $A(k_F,M^*)$ is certainly very small at low densities, 
and actually
it can be still neglected up to a baryon density 
$\rho_B \simeq 3\rho_0$ (see ref.\cite{liu02}).

Then in the density range of interest here we can use, at the leading order,
a much simpler 
form of the symmetry energy, with transparent  
$\delta$-meson effects:
\begin{equation}\label{eq.11}
E_{sym}(\rho_B) = \frac{1}{6} \frac{k_F^2}{E^*_F} + \frac{1}{2}
\left[f_\rho - f_\delta 
\left(\frac{M^*}{E^*_F}\right)^2\right] \rho_B
\end{equation}
We see that, when the $\delta$ is included, the observed $a_4$ value actually
assigns the combination $[f_\rho - f_\delta (\frac{M}{E_F})^2]$
of the $(\rho,\delta)$ coupling constants. If
$f_\delta \not= 0$ we have to increase the $\rho$-coupling (see Fig.1 of
ref.\cite{ku97}). In our calculations we use the value $a_4 = 32  MeV$.

In Table I the $Set~I$ corresponds to $f_\delta=0$. In the $Set~II$
$f_\delta$ is chosen as $2.0fm^2$. Although this value is relatively
well justified, \cite{liu02}, we stress that aim of this work is just
to show the main qualitative new dynamical effects of the $\delta$-meson
coupling.
In order to have the same $a_4$ we must increase the $\rho$-coupling constant
of a factor three, up to $f_\rho=2.8 fm^2$. Now the symmetry energy at 
saturation
density is actually built from the balance of scalar (attractive) and vector
(repulsive) contributions, with the scalar channel becoming weaker 
with increasing
baryon density \cite{liu02}. This is indeed the isovector 
counterpart of the saturation
mechanism occurring in the isoscalar channel for the symmetric nuclear matter. 
From such a scheme we get a further strong fundamental support for the 
introduction
of the $\delta$-coupling in the symmetry energy evaluation.
\begin{figure}[htb]
\epsfysize=6.cm
\centerline{\epsfbox{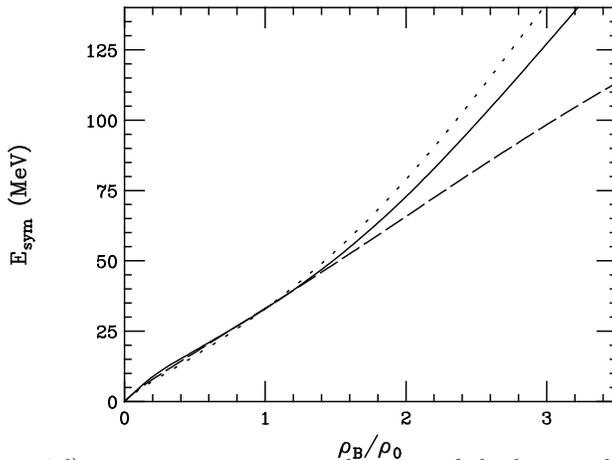}}
\caption{Total (kinetic+potential) symmetry energy as a function of the baryon
density. Long dashed: Hartree ($NLH+\rho$). Dotted: Hartree 
($NLH+\rho+\delta$). Solid: Hartree-Fock ($NLHF$).}
\label{lin1}
\end{figure}
\begin{figure}[htb]
\epsfysize=5.5cm
\centerline{\epsfbox{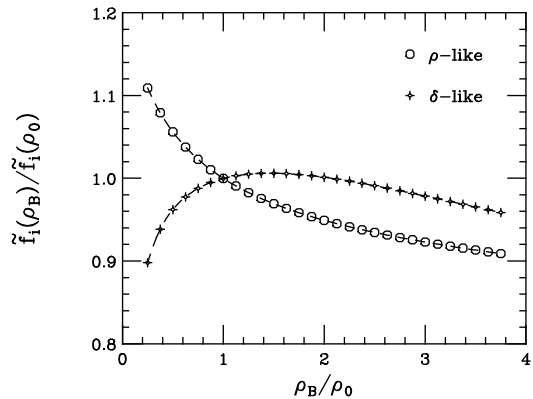}}
\caption{Baryon density variation of the isovector 
effective coupling when the Fock terms are included.}
\label{lin2}
\end{figure}
Details of the calculation can be found in Refs. \cite{gre01,liu02}, here we 
show only the result corresponding to the parametrizations ($Sets~I,~II$ 
and $NLHF$) that we will use to 
investigate the dynamical response.  
In Fig.\ref{lin1} we report the symmetry energy as a function of the 
baryon density in the Hartree case ($\rho$ and $\rho+\delta$) 
and with the Fock terms [($\rho+\delta$) case].
When the $\delta$ ``channel'' is included the behaviour is stiffer 
for the relativistic
mechanism discussed before, see also ref. \cite{liu02}.

When the Fock terms are evaluated the new ``effective'' couplings 
Eqs.(\ref{cincoup})
naturally acquire a density dependence. This is shown in Fig.\ref{lin2} 
for the isovector terms. The decrease of the ``effective'' $\rho$ 
coupling at high density accounts for the slight softening of the 
symmetry energy seen in Fig.\ref{lin1} (from the dotted 
to the solid line).
  
\section{Linear Response Equations}

In this section we study collective oscillations 
that propagate in cold nuclear matter due to the mean field dynamics.
In some sense we follow a relativistic extension of the method
introduced by Landau to study liquid-$^3He$ \cite{landau,abri,baym78} and
recently applied to investigate stable and unstable modes in nuclear matter
\cite{lar98,peth88,bar98}.
 The starting point is
the kinetic transport equation Eq.(\ref{trans}).
 We look for solutions corresponding to small 
oscillations of ${\hat F}(x,p)$ around the equilibrium value. 
Therefore we put 
\begin{equation}
{\hat F}(x,p)={\hat H}(p)+{\hat G}(x,p)
\end{equation}
where $\widehat H(p)$ is the Wigner function at equilibrium 
(see Appendix) and 
$\widehat G(x,p)$ represent its fluctuations.

For the equilibrium state the Wigner function contains  only the isoscalar 
term and the third component of the isovector term. We limit ourselves
to study excited states in these channels. Therefore
we consider isovector density fluctuations 
with $m_T=0$ only, i.e. we do not study processes where a neutron 
converts into a proton or {\it viceversa}. In a linear response scheme  
oscillations in the forementioned channels are decoupled from 
the remaining ones ($m_T=\pm 1$). 
 
In the linear approximation, i.e. neglecting terms of second order in 
$\widehat G(x,p)$, 
the equations for the Wigner functions become 
\begin{eqnarray}\label{elin1}
&&{i\over2}{\partial_\mu}\gamma^\mu{\hat G}_{(1)}(x,p)+
\bigl({\Pi_\mu}-\tilde f_\rho\,b_{\mu}\bigr)\gamma^\mu{\hat G}_{(1)}(x,p)
-M^*_1\,{\hat G}_{(1)}(x,p)=\nonumber\\
&&(1-{i\over2}\Delta)\bigl(\hat{\cal F}(x)+\hat{\cal F}_{3}(x)\bigr)
{\hat H}_{(1)}(p)~, 
\end{eqnarray}
for protons, and
\begin{eqnarray}\label{elin2}
&&{i\over2}{\partial_\mu}\gamma^\mu{\hat G}_{(2)}(x,p)+
\bigl({\Pi_\mu}+\tilde f_\rho\,b_{\mu}\bigr)\gamma^\mu{\hat G}_{(2)}(x,p)
-M^*_2\,{\hat G}_{(2)}(x,p) =\nonumber\\
&&(1-{i\over2}\Delta)\bigl(\hat{\cal F}(x)-\hat{\cal F}_{3}(x)\bigr)
{\hat H}_{(2)}(p)~,
\end{eqnarray}
for neutrons, where $M^*_1=M- \tilde f_\sigma \rho_S - 
\tilde f_\delta\,\rho_{S3}$ and 
$M^*_2=M- \tilde f_\sigma \rho_S + \tilde f_\delta \,\rho_{S3}$. 
The quantities ${\hat {\cal F}}(x)$ and ${\hat {\cal F}}_{3}(x)$  
are the isoscalar and the isovector components of the self--consistent field: 
\begin{eqnarray}
\hat{\cal F}(x)=-8{\tilde f_\sigma}G(x)
+8{\tilde f_\omega}\gamma_\mu{G^\mu}(x) 
-8{{\partial {\tilde f_\sigma}} \over {\partial \rho_S}} \rho_S G(x)
-8{{\partial {\tilde f_\sigma}} \over {\partial j_\mu}} \rho_S  
G^\mu(x)\nonumber\\ 
-8{{\partial {\tilde f_\sigma}} \over {\partial \rho_{S3}}} \rho_S G_3(x)
-8{{\partial {\tilde f_\sigma}} \over {\partial j_{3\mu}}} \rho_S G_3^\mu(x) 
+8{{{\partial {\tilde f_\omega}} \over {\partial \rho_S}}} \gamma_\mu j^\mu G(x)
~,
\end{eqnarray}

\begin{equation}
\hat{\cal F}_{3}(x)=-8 \tilde f_\delta G_{3}(x)
+8 \tilde f_\rho\gamma_\mu G^\mu_{3}(x)
-8{{\partial {\tilde f_\delta}} \over {\partial \rho_S}}~\rho_{S3}~G(x)
+8{{\partial {\tilde f_\rho}} \over {\partial \rho_S}} \gamma_\mu j_3^\mu 
G(x)~.
\end{equation}

The Hartree approximation is recovered by vanishing all the derivaties inside 
the quantities $\hat{\cal F}(x)$ and $\hat{\cal F}_{3}(x)$, except 
${{\partial {\tilde f_S}} \over {\partial \rho_S}}$, since still 
$\tilde f_\sigma=\Phi(\rho_S)/\rho_S$ . 

In order to obtain the equations for the collective oscillations we  
multiply Eqs. (\ref{elin1}) and  (\ref{elin2}) by $\gamma_{\lambda}$. After
performing the traces, and we equate to zero both the real and 
imaginary parts of 
the result \cite{de91,ma94}. Furthermore, by  
Fourier transforming and integrating over four--momentum, we get the 
set of equations  for the scalar and vector 
fluctuation of each species (~$i=1,2$ for proton, neutron respectively): 
%1 equazione
\begin{eqnarray}\label{rpa1}
\sum_{j=1}^2\biggl\{\left[\delta_{i,j}+
\left({\rho_{Si}\over {M^*_i}}-4\,C^{(i)}(k)\right)D^{S}_{ij} + 
4\,{C^\mu}^{(i)}(k){B_\mu^{V}}_{ij}\right]G_{(j)}(k)\nonumber\\ 
+\left[4\,C^{(i)}_\mu(k)D^{V}_{ij} +  
\left({\rho_{Si}\over M^*_i}-4\,C^{(i)}(k)\right){B_\mu^{S}}_{ij}\right]
G_{(j)}^\mu(k)\biggr\}= 0~,
\end{eqnarray}
%2 equazione

\begin{eqnarray}\label{rpa2}
\sum_{j=1}^2\biggl\{\left[\delta_{i,j}g^\lambda_\mu+
\left({\rho_{Si}\over M^*_i}g^\lambda_\mu+4\,{C^{(i)\lambda}_\mu(k)\over 
M^*_i}\right)D^{V}_{ij}
-4\,C^{(i)\lambda}(k){B_\mu^{S}}_{ij}
\right]G_{(j)}^\mu(k)\nonumber\\
-\left[4\,C^{(i)\lambda}(k)D^{S}_{ij} - \left({\rho_{Si}\over M^*_i}
g^\lambda_\mu+4\,{C^{(i)\lambda}_\mu(k)\over M^*_i} 
\right){B_\mu^{V}}_{ij}\right]
G_{(j)}(k)\biggr\}= 0~, 
\end{eqnarray}

with
$$D^{S}_{ij} =  {\tilde f}_{\sigma}+ (-1)^{i+j}\tilde f_{\delta} + 
{{\partial {\tilde f}_\sigma} \over {\partial \rho_S}}\rho_S
-(-1)^{j} {{\partial {\tilde f}_\sigma} \over {\partial \rho_3}}\rho_S 
-(-1)^{i} {\partial {\tilde f_\delta} \over {\partial \rho_S}}\rho_3$$

$$D^{V}_{ij} =  {\tilde f}_{\omega}+ (-1)^{i+j}\tilde f_{\rho} + 
{{\partial {\tilde f}_\omega} \over {\partial \rho_S}}\rho_S
%$$+ {{\partial {\tilde f}} \over {\partial (\rho_j-\rho_z)}}\rho_S 
-(-1)^{i} {{\partial {\tilde f}_\rho} \over {\partial \rho_S}}\rho_3$$

$${B^{V}_\mu}_{ij} =  
{{\partial {\tilde f}_\omega} \over {\partial \rho_S}}j_\mu
-(-1)^i {{\partial {\tilde f}_\rho} \over {\partial \rho_S}}{j_{3\mu}}$$

$${B^{S}_\mu}_{ij} =  
{{\partial {\tilde f}_\sigma} \over {\partial j^\mu}}\rho_S
-(-1)^j {{\partial {\tilde f}_\delta} \over {\partial j^\mu_3}}
\rho_S .$$  
The explicit expressions of the coefficients $C^{(i)}(k)$, 
$C^{(i)}_\lambda(k)$,
 and $C^{(i)}_{\lambda\mu}(k)$, together with some of their properties can be
found in the Appedix.

We notice, however, that the formalism developed in Hartree-Fock approximation
 allows to
achieve the set of equations valid also for an approach to $QHD$ based on 
a dependence on the scalar density of all couplings \cite{flw95}. 
In particular in our case
the coupling functions to the scalar-isoscalar channel depend on all the
isoscalar and isovector densities and currents 
(for details, see \cite{prc_nuovo}),
then it is even more general.

The set of equations developed in the Hartree-Fock approximation
recover the ones correponding to the usual Hartree approximation 
($RMF$). As already mentioned, it
is easly obtained by considering the coupling $\tilde f_i$ to each channel
equal to the coupling constant of the corresponding meson.
The result is an appreciable plainer structure of the equations due to 
the constant value of all couplings, except $\tilde f_\sigma$ \cite{note}.
The Eqs. (\ref{rpa1}) and (\ref{rpa2}) can be reduced to :
\begin{eqnarray}\label{rpa1h}
\sum_{j=1}^2\biggl\{\left[\delta_{i,j}+
\left({\rho_{Si}\over {M^*_i}}-4\,C^{(i)}(k)\right)D^{S}_{ij}\right]G_{(j)}(k)
+\left[4\,C^{(i)}_\mu(k)D^{V}_{ij}\right]
G_{(j)}^\mu(k)\biggr\}= 0~,
\end{eqnarray}

\begin{eqnarray}\label{rpa2h}
\sum_{j=1}^2\biggl\{\left[\delta_{i,j}g^\lambda_\mu+
\left({\rho_{Si}\over M^*_i}g^\lambda_\mu+4\,{C^{(i)\lambda}_\mu(k)\over 
M^*_i}\right)D^{V}_{ij}\right]G_{(j)}^\mu(k)-\left[4\,C^{(i)\lambda}(k)
D^{S}_{ij}\right]
G_{(j)}(k)\biggr\}= 0~, 
\end{eqnarray}

with $D^{S}_{ij} = \frac{d\Phi}{d\rho_S} + (-1)^{i+j} f_{\delta}$, 
$D^{V}_{ij} =  {f}_{\omega}+ (-1)^{i+j} f_{\rho}$.
 
The physics effects appear more trasparent and we will follow the Hartree scheme
in the next sections keeping well in mind that the Fock 
contributions can be easily
included. We expect to have some extra contributions in the various 
interaction channels
without qualitative modifications of the physical response.

The normal collective modes are plane waves, characterized by the wave vector
($k^\mu=(k^0,0,0,\mid \bf{k}\mid))$, they are determined by solving the 
set of homogeneous Eqs.(\ref{rpa1h},\ref{rpa2h}). The solutions 
correspond only to longitudinal waves
and do not depend on
$k^0$ and $\vert {\bf k} \vert$ separetly, but only on the ratio
$$v_s=\frac{k^0}{\vert {\bf k} \vert}~.$$
The sound velocities are given by values of $v_s$ for which the
relevant determinant of the set (Eqs.(\ref{rpa1}, \ref{rpa2})) 
 vanishes, i.e. the dispersion relations. In correspondence the 
neutron/proton structure of the eingenvectors (normal modes) can 
be derived. It should be remarked
that in asymmetric nuclear matter isoscalar and isovector components are
mixed in the normal modes. Here this can be argued by the fact that in each of 
the Eqs. (\ref{rpa1h}, \ref{rpa2h}) both proton/neutron densities  
and currents are appearing.

However we remind that one can still identify  
isovector-like excitations as the modes where neutrons and protons move
out of phase, while isoscalar-like modes are characterized by neutrons
and protons moving in phase \cite{lar98,bar01}.

\section{The Role of Scalar/Vector Fields in the Dynamical Response}
Before showing numerical results for the dynamical response of 
asymmetric nuclear matter,
in various baryon density regions and using the different effective 
interactions,
we would like to analyse in detail the structure of the 
relativistic linear response 
theory in order to clearly pin down the role of each meson coupling.

\subsection*{Isovector Response}

One may expect that once $a_4$ is fixed, the
velocity of sound is also fixed \cite{mat81}. 
On the other hand our results clearly indicate a different
dynamical response with or without the $\delta$-meson channel, for interactions
which give $exactly~the~same~a_4~parameter$, see Figs.(\ref{lin3}, \ref{lin4})
in the following. In order to get a clear understanding of this 
effect we will consider
the case of symmetric nuclear matter in the Hartree scheme, where the dispersion
relations are assuming a transparent analytical form.

For symmetric $NM$ the densities, the effective masses and the coefficients
$C^{(i)}(k)$, $C^{(i)}_{\lambda}(k)$ and $C^{(i)}_{\lambda\mu}(k)$
are equal for protons and neutrons. 
Now it is also possible to decouple the collective modes 
into {\it pure} isoscalar and isovector oscillations, \cite{lar98}. 
 After a straightforward
rearragement of   Eqs. (\ref{rpa1h}), (\ref{rpa2h}), we have 
for the isovector modes:
 
\begin{eqnarray}\label{isovec}
\delta\rho_{3}+\left[{\rho_S\over M^*}-8\,\frac{C_{00}(k)}{M^*}(1-v_s^2)\right]
\, f_\rho\,\delta\rho_{3} -8\,f_\delta\,C_0(k)\,\delta\rho_{S3}=0\nonumber\\
\delta\rho_{S3}+\left[{\rho_S\over M^*}-8\,C(k)\right]f_\delta \, 
\delta \rho_{S3} + 8\,f_\rho\,C_0(k)\,(1-v_s^2)\,\delta\rho_3=0~ . 
\end{eqnarray}   
We stress that the structure is the same for the isoscalar excitations,
of course one has to change the isovector fluctuations with the isoscalar
ones ($\delta \rho_B,~\delta \rho_{S}$), and the coupling constants
of isovector mesons with those of the isoscalar mesons \cite{note}.

Note that in this case to find the zero--sound velocity one has to evaluate
determinant of a $2 \times 2$ matrix (and not a $4\times 4$), hence the 
condition for having a solution can be written as
\begin{equation}\label{det1}
%1+\left[F_\rho(1-v_s^2)- F_\delta\frac{M^{*2}}{E^{*2}_F} 
%\left(\frac{1}{1+f_\delta A(k_F,M^*)}-f_\rho\, \frac{\rho_S}{M^*}\,v_s^2
%\right)
%\right]\varphi(s)=0
1+N_F\left[f_\rho(1-v_s^2)- f_\delta\frac{M^{*2}}{E^{*2}_F} 
\left({1-f_\delta A(k_F,M^*)}-f_\rho\, \frac{\rho_S}{M^*}\,v_s^2\right)
\right]\varphi(s)=0
\end{equation}
here $N_F=\frac{2K_F E^*_F}{\pi^2}$ is the density of states at the
Fermi surface and 
$s\equiv v_s/v_F$. To get Eq.(\ref{det1}) 
we have used the expression for $C(k)$, $C_0(k)$ and $C_{00}(k)$
in terms of the Lindhard function $\varphi(s)$ (see Appendix).
The quantity $A(K_F,M^*)$ is the same integral discussed in 
Eqs.(\ref{eq.9},\ref{eq.10},\ref{eq.11}).

At this point we can make the following approximation
$$v_s^2 \simeq  v_F^2 = \frac{k^2_F}{E^{*2}_F}\,,$$
to evaluate the expression inside the square brackets.
Looking at Fig.s (\ref{lin3},\ref{lin4},\ref{lin7}) this is a good 
approximation within a $3\%$. 
The Eq.(\ref{det1}) asssumes a quite clear form:
\begin{equation}\label{det2}
1+ \frac{6\,E^*_F}{k^2_F}\left[E_{sym}^{pot}-\frac{f_\rho}{2}
\frac{k^2_F}{E^{*2}_F}\left(1-f_\delta\,\frac{M^*}{E^{*2}_F}\,\rho_S\right)
\rho_B\right]\varphi(s)=0\,.
\end{equation}
where the potential part of the symmetry energy explicitly appears in
the dispersion relations, but {\it joined to an important correction term} 
which shows a different $f_\rho,f_\delta$ structure with respect to 
that of $E_{sym}^{pot}$,
Eqs.(\ref{eq.9}, \ref{eq.11}). 
{\it We can easily have interactions with the same
$a_4$ value at normal density but with very different isovector response}.
E.g. when we include the $\delta$ channel we know that we have to increase the
$f_\rho$ coupling in order to have the same $a_4$, see the discussion of the 
Eqs.(\ref{eq.9}, \ref{eq.11}), but now the ``restoring force'' 
(coefficient of the 
Lindhard function in the Eq.(\ref{det2})) will be strongly reduced. 

Equation (\ref{det2}) suggests to define an effective symmetry energy like
\begin{equation}\label{aeff}  
E^{*}_{sym}= E^{pot}_{sym}- \frac{f_\rho}{2}
\frac{k^2_F}{E^{*2}_F}\left(1-f_\delta\, \rho_S\frac{M^*}{E^{*2}_F}\right)
\rho_B.
\end{equation}
which acts as a restoring force for the isovector mode.
We can see that once the symmetry energy is fixed
its effect on the dynamical response depend on the strength of 
each isovector field. In particular we can easily verify that the $f_\delta$
factor inside 
bracket in Eq.(\ref{aeff}) is a second order correction and  the 
 ``leading contribution'' to the reduction  
$\Delta E_{sym}^{*}=E_{sym}^{pot}-E^{*}_{sym}$ is essentially given by 
the coupling of the $\rho$-meson field.
On the other hand we know \cite{gre01} that once the 
symmetry energy at saturation
density $a_4$ is fixed, the change of $f_\rho$ only due to the 
strenght of $f_\delta$.
We have seen from Table I that $f_\rho$ can go from $1~fm^2$, if 
we switch off the $\delta$-channel, to $f_\rho=2.8~fm^2$. 
 In terms of the effective symmetry energy this means (if we consider 
the dynamical
response at $\rho_0$), $\Delta E_{sym}^{*} \sim 4~MeV$ if 
$f_\delta\sim 2.0~ fm^2$. This ``softening'' of the restoring 
force easly accounts for the decrease of the sound velocity 
($v_s/v_{Fn}$ seen in Fig.\ref{lin3})
for symmetric nuclear matter, $\alpha=0$, when we pass 
from $NLH-\rho$ and $NLH-(\rho+\delta$).

\subsection*{Isoscalar Response}
As already remarked, we like to note that for symmetric $NM$ there is a tight 
analogy between the isoscalar and the isovector response in the $RMF$ approach. 
In the isoscalar
degree of freedom the compressibility will play the same role of the symmetry
energy in the dispersion relation equations. 
Also in this case we will have an important correction term 
coming from the interplay of the scalar and vector channel.

The Eq.(\ref{det1}) now becomes \cite{note}:
\begin{equation}\label{dcomp}
1+N_F\left[f_\omega(1-v_s^2)- f_\sigma\frac{M^{*2}}{E^{*2}_F} 
\left({1-f_\sigma A(k_F,M^*)}-f_\omega\, \frac{\rho_S}{M^*}\,v_s^2\right)
\right]\varphi(s)=0
\end{equation}
that can be reduced to the isoscalar equivalent of the Eq.(\ref{det2}):
\begin{equation}\label{dcomp2}
1+ \frac{E^*_F}{3\,k_F^2}\left[K_{NM}^{pot}-9\,f_{\omega}
\frac{k^2_F}{E^{*2}_F}\left(1-f_\sigma\,\frac{M^*}{E^{*2}_F}\,\rho_S\right)
\rho_B\right]\varphi(s)=0\,.
\end{equation}
where the $K_{NM}^{pot}$ is the potential part of the nuclear matter 
compressibility
that in the Hartree scheme has the simple structure \cite{note}(see also Eq.(16)
of ref. \cite{mat81})
\begin{equation}\label{eq.21}
K_{NM}(\rho_B) = \frac{3\,k_F^2}{E^*_F} + 9
\left[f_\omega - f_\sigma 
\left(\frac{M^*}{E^*_F}\right)^2\right] \rho_B 
\equiv K^{kin}_{NM}+K^{pot}_{NM}\,,
\end{equation}

By means of such an analogy, the previous discussion 
can be extended to isoscalar 
oscillations with the role of $E_{sym}$ now ``played'' by the compressibility.
In this case however one always takes into account both the scalar
and vector channel in any $RMF$ models. However the coupling 
costant $f_\omega$ can assume
very different values depending on the required value for effective masses
$M^*_0$. 
This is easy to understand since in the $RMF$ limit the saturation binding
energy has the simple form 
$$E/A(0)=E^*_F+f_\omega\rho_B(0)-M_N$$
where $M_N$ is the bare nucleon mass. So we see that the same saturation values
of $\rho_B$, $E/A$ when decrease $M^*_0$ we have to increase $f_\omega$.
Just to get an idea,
we mention that among the most common used $RMF$ parametrizations, $f_\omega$ 
can go from $3.6\,fm^2$ of $NL2$ \cite{bla93} 
to $10.2\,fm^2$ of $NL3$ \cite{vre97},
just decreasing the effective mass at saturation from $0.82 M_N$ to $0.6M_N$.
In some sense this is obvious because if two $EOS$ have 
different effective masses
even if the compressibility is equal the dynamical behaviour is expected to
be different. This is a very general feature present also
in non-relativistic approaches.

From studies on monopole
resonances in finite nuclei with $RMF$ it seems that a 
higher value of compressibility is required respect to non-relativistic
calculations. Many authors state that this certainly 
demands for a clarification \cite{magi97,magi01}. 
Even if the monopole resonance
is not directly connected to the isoscalar collective mode in nuclear matter,
our discussion nicely suggests to look at the interplay between effective mass
and compressibility.
For example we can estimate by means of Eq.(\ref{dcomp2})that we can have  
a shift between the compressibility and the ``effective compressibility'' of
the order of $\sim 100\, MeV$
among different parametrization with the same $K_{NM}$. Therefore
model with $K \sim 300\, AMeV$ can reproduce the same frequencies
of other models with $K \sim 200\, AMeV$ (and a slightly larger $M^*_0$).

\subsection*{Landau Parameters} 
We would like to briefly discuss the relativistic equations for collective modes
in terms of the Landau parameters. Interesting features will appear from the 
comparison to the non-relativistic analogous case. 
We will focus first on the isovector response, but, as already 
clearly shown before,
 the structure of the results will be absolutely similar in 
the isocalar channel.

The general non--relativistic expression for the isovector modes
can be found in Ref.\cite{abri}:
\begin{equation}\label{nonr1}
1+\left[F^a_0+\frac{F_1^a}{1+1/3\,F^a_1}\,s^2 \right]\varphi(s)=0
\end{equation}
where $F^a_0$ is the ``isovector'' combination of the 
Landau $F_0$ parameters for 
neutrons and protons $F_0^a=F^{nn}_a-F^{np}_0$, that can be expressed in terms
of density variations of the chemical potentials:
\begin{equation}\label{f0}
F^{q q\prime}_0\equiv \frac{\partial \mu_q}{\partial \rho_{q\prime}}
N_q -\delta_{q q\prime}\,,\,\,N_q\equiv \frac{k_{Fq}E^*_{Fq}}{\pi^2}\,,\,\,q=n,p
\end{equation}
$F^a_1$ are the equivalent for the momentum dependent part of the mean field.
In the relativistic approach, for symmetric nuclear matter, we get:
\begin{eqnarray}\label{f1a}
&&F_0^a= F_\rho-F_\delta\,\frac{M^{*2}}{E^{*2}_F}\frac{1}
{1+f_\delta\,A(k_F,M^*)}\nonumber\\
%&&F_1^{s}=-F_\omega\frac{v^2_F}{1+\frac{1}{3}F_\omega\,v^2_F }\nonumber\\
&&F_1^{a}=-F_\rho\frac{v^2_F}{1+ \frac{1}{3}F_\rho\,v^2_F}\,,
\end{eqnarray}
where $F_i=N_F f_i\,(i=\rho,\delta)$ with $N_F=2N_{n,p}$. Note that the $F_1^a$
contribution comes only from the vector coupling. 
By using the expression $E_{sym}^{pot}$ Eq.(\ref{eq.9}), we can write 
Eq.(\ref{nonr1})
in the same form of Eqs.(\ref{det1}, \ref{det2}). The result is a similar 
expression but with the
lack of the term in $f_\delta$ inside the brackets 
in Eq.(\ref{det2}). As said, this is not the leading term, however 
around saturation
density it amounts to about a $10\%$ of the total correction.

 Moreover turning to the analogy with 
isoscalar channel the coupling of the $\sigma$ field is now much larger
 and this purely relativistic contribution could be up to a $20\%$. 
We underline this point because generally the linear
response in $RMF$ is discussed calculating the Landau parameters and then 
using these estimations directly into
the non--relativistic expression for collective modes \cite{mat81,cai01}.

In conclusion from the analysis in terms of the Landau parameters, 
we can describe
the effect of the scalar-vector coupling competition  
previuosly discussed in the
following way. The symmetry energy fixes the $F^a_0$, in fact:
\begin{equation}\label{esim}
E_{sym}=\frac{k_F^2}{6\,E^*_F}(1+F_0^a)
\end{equation}   
but in the dynamical response enters also the $F^a_1$, linked to the momentum
dependence of the mean field, mostly given by the vector meson coupling.
The results are completly analogous in the isoscalar channel, with the
compressibility given by
\begin{equation}
K_{NM}=\frac{3\,k_F^2}{E^*_F}(1+F_0^s)
\end{equation} 
with ``isoscalar'' combination $F^s_0=F^{nn}_0+F^{np}_0$. 
The relativistic forms of
the isoscalar Landau parameters are exactly the same as in Eq.(\ref{f1a}), just
substituting the $\delta,\rho$ coupling constants with the $\sigma,\omega$ ones
\cite{note}.

\section{Isovector Collective Modes in Asymmetric Nuclear Matter}
In this section we discuss results for the isovector collective oscillations
which are driven by the symmetry energy terms of the nuclear $EOS$.
The aim is mainly to investigate the effect of the scalar-isovector channel.
This is normally not included in studying the isovector modes and 
in general the properties of symmetric matter in a relativistic approach,
while it should be naturally present on the basis of the analysis shown in the
previous section (and in ref.\cite{liu02} for equilibrium properties).
Moreover we stress again that Hartree-Fock scheme embodies 
in any case the presence
of a scalar-isovector channel, {\it even without the 
inclusion of the $\delta$-meson field}\cite{gre01}.

We will first show results obtained in the Hartree scheme ($NLH$) including
either both the isovector $\rho$ and $\delta$ mesons or only the $\rho$ meson.
Even if the Hartree approximation has a simpler structure, it contains 
all the physical
effects we want to point out. Finally from the complete 
Hartree-Fock ($NLHF$) calculations we will confirm the dynamical contribution
of the scalar isovector channel.

For $NLH$ calculations  we use the parametrizations of Table I, Set I ($\rho$)
and Set II ($\rho+\delta$). In the Hartree-Fock case the coupling 
constant $f_\delta$ is adjusted to the value 
$\tilde f_\delta(\rho_0)=2.0 fm^2$ of the $NLHF$ model
Eq.(\ref{cincoup}). We note that this value is smaller that the prediction of
recent Dirac-Brueckner-Hartree-Fock calculations \cite{hole01}, therefore the
effects due to the $\delta$-channel presented in the following could be even
underestimated in the $NLHF$ case.

\subsection*{Hartree Results}

\begin{figure}[htb]
\epsfysize=7.0cm
\centerline{\epsfbox{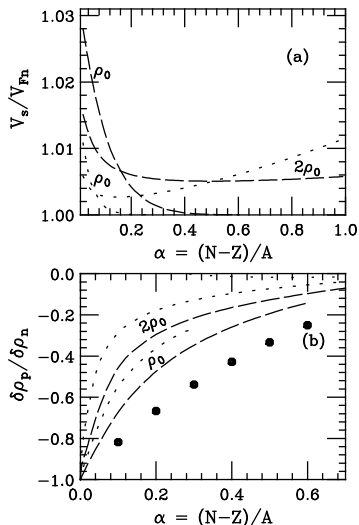}}
\caption{Isovector-like modes: (a) Ratio of zero sound velocities to the neutron
Fermi velocity $V_{Fn}$ as a function of the asymmetry parameter $\alpha$ for
two values of baryon density. Long dashed line:$NLH-\rho$; 
Dotted line:$NLH-(\delta+\rho)$. (b) Corresponding ratios of proton 
and neutron amplitudes. All lines
are labelled with the baryon density, $\rho_0=0.16 fm^{-3}$. The full circles
in panel (b) represent the trivial behaviour of $-(\rho_p/\rho_n)$
vs. $\alpha$.}
\label{lin3}
\end{figure}

Let us start by considering isovector-like excitations. 
In Fig.\ref{lin3}a we show the sound velocities in the Hartree approximation, 
as a function of the asymmetry parameter $\alpha$ for different baryon
densities. We actually plot the sound velocities in units of the neutron Fermi 
velocities. This is physically convenient: when the ratio is approaching
1.00 we can expect that this ``zero'' sound will not propagate due
to the strong coupling to the ``chaotic'' single particle motions 
(``Landau damping''). This quantity then will also directly give a measure
of the ``robustness'' of the collective mode we are considering.

Dotted lines refer
to calculations including ($\rho+\delta$) mesons, long-dashed lines
correspond to the case with only the $\rho$ meson.  Calculations are
performed at $\rho_B = \rho_0$ and $\rho_B = 2~\rho_0$. 
We stress that the results of the two calculations differ already
at zero asymmetry, $\alpha=0$. 
At normal density ($\rho_0$ curves),
in spite of the 
fact that the symmetry energy coefficient, $a_4=E_{sym}(\rho_0)$, 
is exactly the same in the two cases, significant differences are observed in
the response of the system. 
From Fig.\ref{lin3}(a) we can expect a reduction of the 
frequency for the bulk isovector dipole mode in stable nuclei when the 
scalar isovector channel ($\delta-like$) is present. Moreover we note that, in
the $NLH-\rho$ case, the exicitation of isovector modes persists up to higher
asymmetries at saturation density.
These are non-trivial features, related to the different way scalar and vector
fields are entering in the dynamical response of the nuclear system. Such
behaviours are therefore present in both collective responses, isoscalar and
isovector. We have devoted the whole previous section (Sec.IV) to 
a complete discussion
of this effect. 

Differences are observed even at 
$\rho_B = 2\rho_0$, where however also the symmetry energy is 
different. A larger $E_{sym}$ is obtained in the case 
including the $\delta$ meson (see Fig.\ref{lin1}) and this leads to 
a compensentation of the 
effect observed at normal nuclear density. In particular, at higher
asymmetries $\alpha$ the collective exicitation becomes more robust
for $ NLH-(\rho+\delta)$. 
Differences are observed also in the "chemical" structure of the mode, 
represented by the ratio $\delta \rho_p/\delta \rho_n$, plotted in 
Fig.\ref{lin3}(b).
The ratio of the out of phase $n,p$ oscillations is not following the ratio
of the $n,p$ densities for a fixed asymmetry, given by the full circles in the
figure. We systematically see a larger amplitude of the neutron oscillations.
The effect is more pronouced when the $\delta$ (scalar-isovector) channel
is present (dotted lines).

\subsection*{Hartree-Fock Results}
 We have also performed the calculation in the more general case of the 
Hartree-Fock approximation, $NLHF$, whose formalism has been presented 
before, Eqs.(\ref{rpa1}, \ref{rpa2}). We have
fitted the same properties of symmetric $NM$ at the saturation density
as for Hartree case, $NLH$. In particular at $\rho_0$ the value of 
the isovector coupling is fixed in order get the same 
symmetry energy (the $a_4$ parameter)
of the $NLH-(\rho+\delta)$ case.
\begin{figure}[htb]
\epsfysize=7.0cm
\centerline{\epsfbox{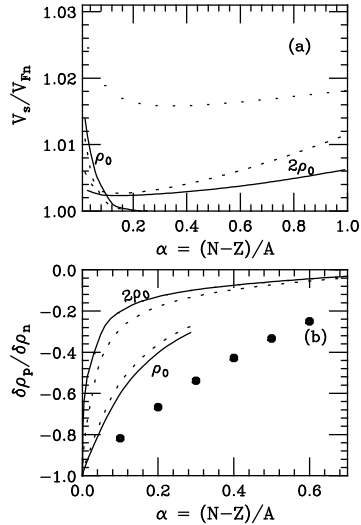}}
\caption{The same as Fig.3 but for $NLHF$ (solid lines) and $NLH-(\delta+\rho)$
(dotted lines).}
\label{lin4}
\end{figure}
In Fig.\ref{lin4} we can see that quite similar results are 
obtained in Hartree-Fock calculations, with respect to the 
Hartree results including $\rho$ and
$\delta$ mesons, especially at the normal density. This can be 
understood by considering that in Hartree-Fock calculation the 
effective density dependent 
couplings associated with the isovector channels are tuned in such a way to 
roughly reproduce, at normal density, the values of the coupling constants 
$f_\rho$ and $f_\delta$ of the Hartree scheme: then not only $a_4$ is the same
but also its internal structure. Since such a tuning can be done only at 
a given density value, some differences are observed at $\rho_B=2\,\rho_0$, due
to the density dependence of the effective coupling constants of the
$NLHF$ scheme. Therefore the greater value of sound velocity in the Hartree
case can be linked to the greater value of $E_{sym}$, see Fig.\ref{lin1}.

\subsection*{Disappearance of the Isovector Modes}
\begin{figure}[htb]
\epsfysize=7.0cm
\centerline{\epsfbox{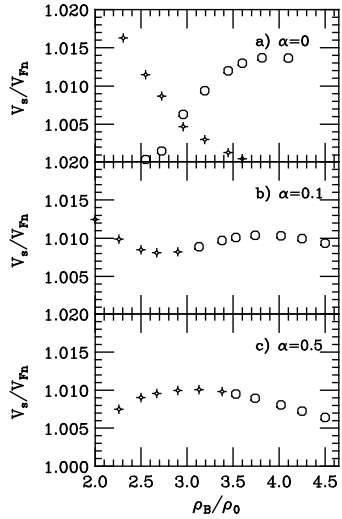}}
\caption{Sound phase velocities of the propagating collective 
mode vs. the baryon density ($NLH+\rho$ case). Crosses: isovector-like. 
Open circles:
isoscalar-like. (a): symmetric matter. (b): asymmetric matter, $\alpha=0.1$.
(c): asymmetric matter, $\alpha=0.5$.}
\label{lin5}
\end{figure}

\begin{figure}[htb]
\epsfysize=4.7cm
\centerline{\epsfbox{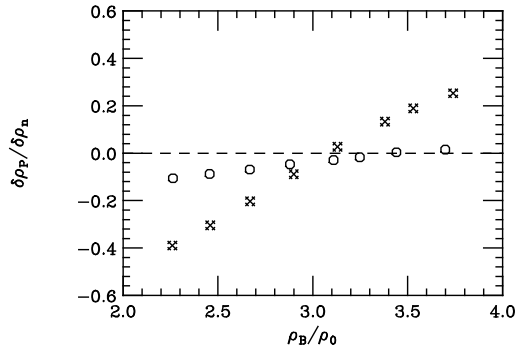}}
\caption{Ratio of protons and neutron amplitudes in the propagating mode, for
different asymmetries, as a function of the baryon density around 
the $\rho_B^{cross}$. Crosses: $\alpha=0.1$, Fig.5b. Open circles: 
$\alpha=0.5$, Fig.5c.}
\label{lin6}
\end{figure}

For asymmetric matter we have found that, in all the calculation schemes, with
increasing baryon density the isovector modes disappear: we call such densities
$\rho_B^{cross}$. E.g. from Figs.\ref{lin3}(b),\ref{lin4}(b) we see that 
the ratio $\delta \rho_p/\delta \rho_n$ tends very quickly to zero with 
increasing
baryon density, almost for all asymmetries. Around this transition 
density we expect to have an almost $pure~ neutron~wave$ propagation of 
the sound.
 Here we show the results of the $NLH+\rho$ case, see Figs.\ref{lin5} and
\ref{lin6}, but the effect is clearly present in all the models.

For symmetric matter we have a real crossing of the two phase velocities, 
isoscalar and isovector, as shown in Fig.\ref{lin5}(a). Above $\rho_B^{cross}$
the isoscalar mode is the most robust.

For asymmetric matter we observe a transition in the structure of 
the propagating normal mode, from isovector-like to isoscalar-like, 
Fig.\ref{lin5}(b,c).
Similar effects have been seen in a non-relativistic picture \cite{lar98}.

For a given asymmetry $\alpha$ the value of $\rho_B^{cross}$ is different
for the three models considered, as can be argued by the behaviour of 
$\delta \rho_p/\delta \rho_n$ at $2\rho_0$ in 
Figs.\ref{lin3}(b), \ref{lin4}(b).
E.g. for $\alpha=0.1\,~NLHF$ has the lower value 
($\rho_B^{cross}\simeq 2.4 \rho_0 $), while $NLH-\rho$ has the higher one 
($\rho_B^{cross}\simeq 3.0 \rho_0$).
This is again related to the reduction of the isovector restoring force when the
scalar-isovector channel ($\delta$-like) is present, see Sec.IV.

From Fig.\ref{lin6} we see that the proton component of the propagating sound
is quite small in a relatively wide region around the ``transition'' 
baryon density, a feature becoming more relevant with increasing asymmetry, 
see the open 
circle line. This is quite interesting since it could open the possibility
of an experimental observation of the $neutron~wave$ effect.
 
\section{Isoscalar Collective Modes in Asymmetric Nuclear Matter}

So far we have focussed our discussion on the isovector-like response of 
the asymmetric nuclear
matter. However it is well known that in asymmetric nuclear matter can
exist also isoscalar-like modes, see \cite{lar98,bar01} and ref.s therein.

\subsection*{Exotic High Baryon density modes}
From the previous analysis we have seen the isoscalar-like excitations 
to become
dominant at high baryon density, above the $\rho_B^{cross}$ introduced before.

\begin{figure}[htb]
\epsfysize=7.0cm
\centerline{\epsfbox{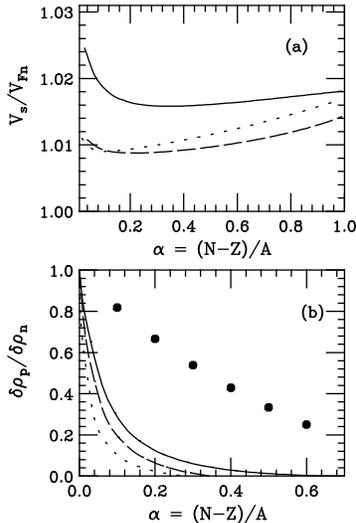}}
\caption{The same of Fig.3 for isoscalar-like modes, at $\rho_B=3.5\rho_0$.
Solid line: $NLHF$. Long Dashed line: $NLH-\rho$. Dotted line: $NLH-(\rho+\delta)$. The full circles in panel (b)
represent the behaviour of $\rho_p/\rho_n$ vs. $\alpha$.}
\label{lin7}
\end{figure}

Some results are shown in Fig.\ref{lin7}.
It should be noticed that 
the frequency of the isoscalar-like modes is essentially related to the 
compressibility of the system at the considered density. 
In Fig.\ref{lin7}(a) we display the sound velocity obtained in Hartree and 
Hartree-Fock calculations at $\rho_B = 3.5~\rho_0$, as a function 
of the asymmetry $\alpha$.   
The differences observed among calculations performed within the Hartree or
Hartree-Fock scheme are due to a different behaviour of   
the associated equation of state at high density.  
%Also the $I$ dependence is different in
%the three cases. 
At $\alpha=0$ the two Hartree models have exactly the
same isoscalar mean fields, but for asymmetric nuclear matter
the different behaviour of the symmetry energy
leads to a different compressibility. The case 
$NLH-(\rho+\delta)$ which has the stiffer $E_{sym}$ 
(resulting in a greater incompressibility for $\alpha\,>\,0$) with respect to
$ NLH-(\rho)$ shows 
also a greater increase of $v_s/v_{Fn}$ with density.
Instead, $NLHF$ even if it has the same compressibility $K_{NM}$ at 
saturation density,
shows a different $v_s$. This should be due to the density dependence of the
coupling function arising from exchange terms which leads to different
values of $K_{NM}$ out of $\rho_0$ (even for $\alpha=0$).

Some differences are observed also in the chemical composition
of the mode (Fig.\ref{lin7}(b)). The black spots show the behaviour 
of $\rho_p/\rho_n$ vs. $\alpha$. Note the $pure~neutron~wave$ structure 
of the propagating
sound, since the oscillations of
protons appear strongly damped ($\delta\rho_p/\delta\rho_n\ll \rho_p/\rho_n$);
unfortunately this is an effect not experimentally accessible (at present), see
also the discussion at the end of the previous Section. 

Before closing this discussion we have to remark that the isoscalar-like modes
at high baryon density are vanishing if the nuclear $EOS$ becomes softer. This
is indeed the results of two recent models, ref.s \cite{cai01,wolter}, where
the nuclear compressibility is decreasing at high baryon density for a 
reduction of the isoscalar  vector channel contribution. In \cite{cai01} 
this is due to
self-interacting high order terms for the $\omega$ meson, while in 
ref.\cite{wolter} to a
reduced $f_\omega$ coupling with increasing baryon density.

Finally we note that all causality violation problems (superluminal sound
velocities) observed in the non relativistic results at high baryon density, see
\cite{mat81} and Fig.3c in ref.\cite{lar98}, are completely absent 
in the relativistic approach, see the high density trends in Fig.\ref{lin5}.

\subsection*{Isospin Distillation in Dilute Matter}

\begin{figure}[htb]
\epsfysize=7.0cm
\centerline{\epsfbox{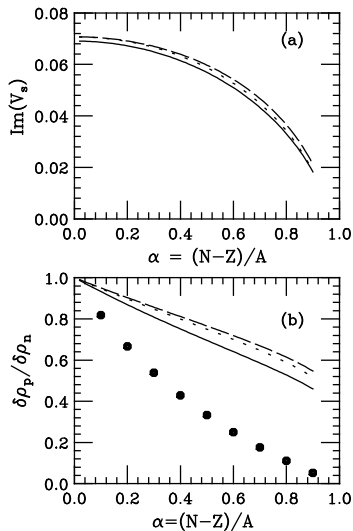}}
\caption{Isocalar--like unstable modes at $\rho_B=0.4\,\rho_0$ : 
Imaginary sound velocity (a) and ratio of proton and neutron 
amplitudes (b)  as a function of the
asymmetry $\alpha$. Solid line: $NLHF$; Dotted line: 
$NLH-(\rho+\delta)$; Long Dashed line: $NLH-\rho$.The full circles in panel (b)
represent the behaviour of $\delta \rho_p/\delta \rho_n$ vs. $\alpha$.} \label{lin8}
\end{figure}

We have also investigated the response of the system in the region of spinodal
instability associated with the liquid--gas phase transition,
 which occurs at low densities. It is known that in this region
an isoscalar unstable mode can be found, with 
immaginary sound velocity, that gives rise to an exponential growth of the
fluctuations. The latter can represent a dynamical mechanism for
the multi--fragmentation
process observed in heavy--ion collisions. 
We have found this kind of solution in the present approach. 
In Fig.\ref{lin8} 
we show the ratio $\delta\rho_p/\delta\rho_n$ as function of
the initial asymmetry for such a collective mode. For all the 
interactions this ratio is different from  
the corresponding $\rho_p/\rho_n$ of the initial asymmetry $\alpha$. This is 
exactly the chemical effect associated with the new instabilities in
dilute asymmetric matter \cite{mu95,bar01}.
In particular  
it is found that, when isoscalar-like modes become unstable,
the ratio $\delta\rho_p/\delta\rho_n$ 
becomes $larger$ that the ratio  $\rho_p/\rho_n$ (at variance with 
the stable modes at high densities, see Fig.\ref{lin7}). Hence proton
oscillations are relatively larger than neutron oscillations leading
to a more symmetric liquid phase and to a more neutron rich gas phase, 
during the disassembly of the system. 
This is the so--called isospin distillation effect in fragmentation, and
signatures of this effect could be searched by looking at the ratio
$N/Z$ of fragments produced in dissipative heavy ion collisions 
\cite{xu00,dit01}.
 
We note here that in dilute asymmetric $NM$ we can distinguish two regions 
of instability, mechanical (cluster formation) and chemical (component
separation). There is however no discontinuity in the structure
of the unstable modes which are developing. For all realistic
effective nuclear interactions (relativistic and non) the nature of the 
unstable normal modes at low densities is always {\it isoscalar-like},
i.e. with neutrons and protons oscillating in phase, although
with
a distillation effect discussed before, see ref.\cite{bar01}
for a fully detailed study of this important property
of asymmetric nuclear matter.

In Fig.\ref{lin8} we observe that Hartree results 
(with and the without $\delta$
meson) are very similar and, indeed, at low density the symmetry
energy behaviour is nearly the same in the two cases. 
On the other hand, differences are observed with respect to the 
Hartree-Fock case. In fact, the $NLHF$ symmetry energy presents a softer 
behaviour (around $\rho_B=0.4\rho_0$), that leads to a smaller
distillation effect. 
We remark that the equality between $NLH$ with and without $\delta$ is
in agreement with the analysis
in terms of the generalized Landau parameters associated to 
normal modes developed in Ref. \cite{liu02}. We can conclude
that there are essentially no effects of the scalar isovector channel
on isospin distillation in the spinodal decomposition.

\section{Conclusions and Outlook}
We have developed a linear response theory starting from a 
relativistic kinetic equations deduced within a Quantum-Hadro-Dynamics 
effective field picture of the hadronic phase of nuclear matter. In the 
asymmetric case we consider as the
main dynamical degrees of freedom the nucleon fields coupled to 
the $isoscalar$, scalar $\sigma$ and $\omega$, and to the $isovector$, 
scalar $\delta$ 
and vector $\rho$, mesons.

Using the Landau procedure we derive the dispersion relations which
give the sound phase velocity and the internal structure of the normal 
collective modes, stable and unstable.
We have focussed our attention on the effect of the isovector mesons on the 
collective response of asymmetric (neutron-rich) matter. In order to better 
understand the dynamical role of the different mesons, the results are obtained
in the Hartree approximation, which has a simpler and more trasparent form.
The contribution of Fock terms is also discussed.

We have singled out some qualitative new effects of the $\delta$-meson-like 
channel on the dynamical response of $ANM$. Essentially, our investigation 
indicates
that even if the symmetry energy is fixed, the dynamical response is affected by
its internal structure, i.e. the presence or not of an isovector-scalar field.
This is implemented by the explicit introduction of an effective $\delta$-meson
 and/or by the Fock term contributions. The interest comes from
the fact that both mechanisms are absent in the usual relativistic $RPA$
calculations for finite nuclei.

It is important to stress that the same interplay between scalar 
($\sigma$-meson) and vector ($\omega$-meson) contributions can be seen in the
dynamical isoscalar response. In general we clearly show a close analogy 
in the structure of the linear response equations:
\begin{itemize}
\item Same form of the dispersion relations, cfr. Eq.(\ref{det1}) 
and Eq.(\ref{dcomp}).
\item Parallel role of $E_{sym}^{pot}$ and $K^{pot}_{NM}$ in the determination
of the restoring force, Eqs.(\ref{det2}) and (\ref{eq.21}).
\item Parallel structure of the corrections due to the scalar-vector meson
competition, Eqs.(\ref{det2}) and (\ref{eq.21}).
\end{itemize} 

This appears to be a beautiful ``mirror'' structure of the 
relativistic approach that strongly support the introduction of a 
$\delta$-meson-like coupling 
in the isovector channel. We like to remind that the same ``mirror'' structure
of the relativistic picture has been recently stressed in ref.\cite{liu02} for
equilibrium properties, saturation binding and symmetry energy, the 
$a_1$ and $a_4$ parameters of the Weiszaecker mass formula.

The relativistic dispersion relations have been compared with the 
non-relativistic ones of the Landau Fermi Liquid theory expressed in 
terms of Landau
parameters evaluated in a relativistic scheme \cite{cai01}. The 
corrections appear to be not negligible, particularly for the 
isoscalar response.

From the numerical results on the collective response of $ANM$ some 
general features are qualitatively present in all the effective 
interactions in the isovector channel:
\begin{itemize}

\item In asymmetric matter we have a mixing of pure isoscalar and pure isovector
oscillations which leads to $chemical$ effect on the structure of the 
propagating collective mode:
the ratio of the neutron/proton density oscillations 
$\delta \rho_n/\delta \rho_p$ is different from the initial $\rho_n/\rho_p$ 
of the matter at equilibrium.
However we can still classify the nature of the excited collective motions as
$isoscalar-like$ (when neutron and protons are oscillating in phase) and 
$isovector-like$ (out of phase). To note that similar effects can be obtained
also using non-relativistic effective forces \cite{lar98,bar01}.

\item For a given asymmetry the isovector-like mode is the most robust at 
low baryon density, always showing a larger neutron component in the oscillations.
With increasing baryon density we observe a smooth transition, at a $\rho_B^{cross} \simeq 2-3\rho_0$, to an isoscalar-like branch, still with a dominant
$\delta\rho_n$. In the region of the transition we predict a propagation of
almost $pure~neutron~waves$. For relatively large asymmetries
($\alpha\equiv\frac{N-Z}{N+Z}=0.5, N=3Z$) this behaviour is present in a wide
interval of densities around $\rho_B^{cross}$. All that seems to suggest 
the possibility of an experimental observation of related effects in 
intermediate energy heavy ion collisions with exotic beams. If the 
compressibility of nuclear
matter is decreasing at high baryon density also these exotic isoscalar-like
mode will disappear. This could be a nice signature of the softening of nuclear
$EOS$ at high densities.

\item The isoscalar-like motions become unstable at sub-saturation densities
still with a strong chemical effect, now in the opposite direction with 
respect to the one discussed before, present in the stable high density modes.
Now the unstable oscillation is more proton-rich, eventually leading to 
the formation of more symmetric clusters vs. a very neutron-rich gas phase. 
This is the
$neutron~distillation$ effect \cite{mu95,lar98,bar98,bar01,xu00,dit01}, a new
important feature of the liquid-gas phase transition in asymmetric nuclear
systems.
\end{itemize}

\subsection{Acknowledgements}
We warmly thank Hermann H. Wolter and  Stephan Typel for several pleasant
and stimulating discussions.

\section{Appendix}
%\chapter{Fourier Transform Coefficients}
%%%%%%%APPENDIX%%%%%%%%%%%%%
The Wigner matrix $\hat H(p)$ for matter at equilibrium saturated in spin
has the following form:
$$\hat H(p)=H(p)+\gamma_\mu H^\mu(p) .$$
>From the kinetic equation one obtain the relation for nuclear matter at equilibrium between the scalar and vector parts:
$$H^\mu(p)=\frac{p^*_\mu}{M^*}H^\mu(p)$$
where the zero component of the vector part is proportional to the Fermi Dirac 
distribution function:
$$H_0^{(i)}(p)=\frac{1}{4}\frac{1}{(2\pi)^3} \Theta(E^*_{Fi}-E^*_{ki})
\delta(p^{*(i)}_0-E_{ki}^*) $$
where $i=n,p$ and $E^*_{ki}=(k^2+M^{*2}_i)^{1/2}$.

The coefficients $C^{(i)}(k)$, $C^{(i)}_{\lambda}(k)$ 
and $C^{(i)}_{\lambda\mu}(k)$ introduced in Sec. II are given 
by the integrals 
$$C^{(i)}(k)=M^*_i\,\int d^4p{H^{\prime\lambda}_{(i)}(p)k_\lambda\over 
p_\rho^{*i} k^\rho}~,\eqno(A1a)$$
$$C^{(i)}_\lambda (k)=\int d^4p{H^{\prime\mu}_{(i)}(p)k_\mu
p^{*i}_\lambda\over p^{*i}_\rho k^\rho}~,\eqno(A1b)$$
$$C^{(i)}_{\lambda\mu} (k)=\int d^4p{H^{\prime\nu}_{(i)}(p)k_\nu
p^{*i}_\lambda p^{*i}_\mu\over p^{*i}_\rho k^\rho}~,\eqno(A1c)$$
where $H^{\prime\lambda}_{(i)}(p)=\partial^\lambda_{(p)}H_{(i)}(p)$. The 
index $i$ specifies the kind of nucleon: $i=1$ for protons 
and $i=2$ for neutrons.
The frequency $k^0$ includes an imaginary part $i\epsilon$ with $\epsilon$ 
positive infinitesimal.
\par
By using the definitions (A1) it can be easily checked that 
$$k^\lambda C^{(i)}_\lambda(k)=0
,~~~k^\lambda C^{(i)}_{\lambda\mu}
(k)=-k_\mu{\rho_{Si}\over 4}~,$$ 
$$\sum_\mu C^{(i)\mu}_\mu(k)=M^*_i 
C^{(i)}(k)-{1\over 2}\rho_{Si}~.\eqno(A2)$$
In order to be more specific we choose the $z$ axis in the direction of 
the wave vector ${\bf k}$. As a consequence, the following coefficients 
identically vanish:\hfill\break
\vskip 0.05cm 
~~~~~$C^{(i)}_1(k),~~ C^{(i)}_2(k),~~ C^{(i)}_{10}(k),~~ C^{(i)}_{20}(k)$,~~~~~
 and ~~~~~~~$C^{(i)}_{lm}(k)$\hfill\break
\vskip 0.05cm
\noindent
for $l\ne m$ ($l$ and $m$ are space indices ). In addition, for symmetry 
reasons,\hfill\break
\vskip 0.05cm
~~~~~$C^{(i)}_{11}(k)=C^{(i)}_{22}(k)$~.\hfill\break
\par
The integrals in Eqs. (A1) can be evaluated analytically. They give
$$C^{(i)}(k)=-{1\over 2}\,{1\over M^*_i}\rho_{Si}+{3\over 4}
\frac{\rho_{Bi}}{E^*_{Fi}} 
-{1\over 4}N_i\frac{M^{*^2}_i}{E^*_{F_i}}\,\varphi(s_i)~,\eqno(A3a)$$
$$C^{(i)}_0=+{1\over 4}N_i\,\varphi(s_i)~,\eqno(A3b)$$
$$C^{(i)}_{00}=-{1\over 4}\,\rho_{Si}+
{1\over 4}N_i\,M^{*}_i\,\varphi(s_i)~,\eqno(A3c)$$
$$C^{(i)}_{11}(k)={1\over 4}\,\rho_{Si}-{3\over 8}\,\frac{M^*_i}{E^*_{Fi}} 
\rho_{Bi}+{3\over 8}\frac{M^*_i}{E^*_{Fi}}(s_i^2-1)\,
\varphi(s_i)~,\eqno(A3d)$$
where $v_{F_i}$ is the Fermi velocity, $s_i=k^0/(v_{F_i}|{\bf k}|)$, 
$N_i$ are the density of states at Fermi surface and  
$$\varphi(s_i)=1-{s_i\over 2}ln\left|{s_i+1\over s_i-1}\right|+{i\over 2}
\,\pi s_i\,\theta(1-s_i)~$$
is the Lindhard function.
The remaining coefficients $C^{(i)}_3(k)$, $C^{(i)}_{03}(k)$ 
and $C^{(i)}_{33}(k)$ can be evaluated by means of the relations (A2).
  
%%%%%%%%%%%%%%%%%%%%%%%%%%%%%%%%%%%%%%%%%%%%%%%%%%%%%%%%%%%%%%%%%%%%%
%%%%%%%%%%%%%%%%%%%%%%%%%%%%%%%%%%%%%


\begin{references}
\bibitem{wal74} J.D.Walecka, Ann.Phys. (N.Y.) {\bf 83}, 491 (1974).
\bibitem{Se86}B. D. Serot and J. D. Walecka, in {\it Advances in Nuclear 
Physics}, edited by J. W. Negele and E. Vogt (~Plenum, New York, 1986~) 
Vol. 16. 
\bibitem{Se97}B. D. Serot and J. D. Walecka, 
Int. J. Mod. Phys. E {\bf 6}, 515 (1997). 
\bibitem{snr94} M.M. Sharma, M.A. Nagarajan and P. Ring, Ann. Phys. (NY)
{\bf 231}, 110 (1994) ,\\
G.A. Lalazissis, M.M. Sharma, P. Ring and Y.K. Gambhir,
Nucl. Phys. {\bf A608}, 202 (1996).
\bibitem{ri96}P. Ring, Prog. Part. Nucl. Phys. {\bf 37}, 193 (1996).
\bibitem{tywo99} S. Typel and H.H. Wolter, Nucl. Phys. {\bf A656}, 331 (1999).
\bibitem{mu95} H.M\"uller and B.D.Serot, Phys.Rev. {\bf C52}, 2072 (1995).
\bibitem{bla93}B. Blattel, V. Koch and U. Mosel, Rep. Prog. Phys. {\bf 56}, 1
 (1993).
\bibitem{koli96}C.M. Ko and G.Q. Li, J. of Phys. {\bf G22}, 1673 (1996).
\bibitem{de91}A. Dellafiore and F. Matera, Phys. Rev. C {\bf 44},  
2456 (1991). 
\bibitem{ma94}F. Matera and V. Yu. Denisov, Phys. Rev. C {\bf 49},  
2816 (1994).
\bibitem{vre97}D. Vretenar, G.A. Lalazissis, R. Behnsch, W. Poeschl and 
P. Ring, Nucl.Phys. {\bf A621}, 853 (1997).
\bibitem{magi97} Z.Y. Ma , N. Van Giai, H. Toki and M. L'Huillier,
Phys. Rev. {\bf C 55}, 2385 (1997).
\bibitem{magi01}Z.Y. Ma , N. Van Giai, A. Wandelt, D. Vretenar and P. Ring,
Nucl. Phys. {\bf A686}, 173 (2001) and ref.s there in. 
\bibitem{rpa00} C.J. Horowitz and K Wehrberger, Nucl. Phys. {\bf A531}
, 665 (1991).
\bibitem{mat81} T.Matsui, Nucl. Phys. {\bf 365}, 365 (1981).
\bibitem{song} C. Song, Phys. Rep. {\bf 347}, 289 (2001).
\bibitem{cai01}J.C. Caillon, P. Gabinski and J. Labarsoque, Nucl. Phys. {\bf A696}, 623 (2001).
\bibitem{liu02} B.Liu, V.Greco, V.Baran, M.Colonna and M.Di Toro,
 Phys.Rev. {\bf C65}, 045201 (2002).
\bibitem{lar98} M.Colonna, M.Di Toro and A.B.Larionov
Phys. Lett. {\bf 428B}, 1 (1998).
\bibitem{bo89}J. Boguta and C.E. Price, Nucl. Phys. {\bf A505}, 123 (1989). 
\bibitem{hor83}C.J. Horowitz and B.D. Serot, Nucl.Phys. {\bf A399}, 529 (1983).
\bibitem{bou87}A. Bouyssy, J.F. Mathiot, Nguyen Van Giai, and S. Marcos, 
Phys. Rev. {\bf C36}, 380 (1987). 
\bibitem{prc_nuovo} V. Greco, F. Matera, M. Colonna, M. Di Toro and G. Fabbri,
Phys. Rev. {\bf C 63}, 035202 (2001).
\bibitem{gre01} V.Greco, M.Colonna, M.DiToro, G.Fabbri and F.Matera,
 Phys.Rev. {\bf C64}, 045203 (2001).
\bibitem{degr80} S.R.de Groot, W.A. van Lee and Ch.G. van Weert,
{\it Relativistic Kinetic Theory}, North-Holland Amsterdam 1980.
\bibitem{hak78} R.Hakim, Nuovo Cimento {\bf 6},1 (1978).
\bibitem{lari97} G.A.Lalazissis, J.Konig and P.Ring, Phys.Rev. {\bf C55},
 540 (1997).
\bibitem{ku97} S.Kubis and M.Kutschera, Phys.Lett. {\bf B399}, 191 (1997).
\bibitem{landau} L.D. Landau and E.M. Lifshitz, {\it Statistical Physics},
 Pergamon Press,Oxford 1980, pag. 288. 
\bibitem{abri}A.A. Abrikosov and I.M. Khalatnikov, Rep. Prog. Phys. 
{\bf 22}, 329 (1959).
\bibitem{baym78} G. Baym and C. J. Pethick in {\it The physics of Liquid 
and Solid Helium} edited by K.H. Bennemann and J.B. Ketterson, Vol 2,
(Wiley, New-York, 1978), p.1.
\bibitem{peth88} C.J. Pethick and D.G. Ravenhall, Ann. Phys. (N.Y.)
{\bf 183}, 131 (1988).
\bibitem{bar98}V. Baran, M. Colonna,M. Di Toro and A. B. Larionov, Nucl. Phys.
{\bf A632}, 287 (1998) .
\bibitem{flw95} Ch. Fuchs, H. Lenske and H.H. Wolter,
Phys. Rev. {\bf C52}, 3043 (1995).
\bibitem{note}We remind that if $\phi$ field has self-interaction terms instead
of $f_\sigma$ one has to put $\frac{d\phi}{d\rho_S}$.
\bibitem{bar01} V.Baran, M. Colonna, M. DiToro and V.Greco,
Phys. Rev. Lett. {\bf 86}, 4492 (2001).
\bibitem{hole01} F.Hofmann, C.M. Keil and H.Lenske, Phys.Rev. {\bf C64},
 034314 (2001).
\bibitem{wolter}S. Typel and H.H. Wolter, private communication and
{\it Collective Modes of Nuclear Matter in Relativistic Mean Field Theories}
Preprint LMU-Physik Section, Munich 2002.

\bibitem{xu00} H.S.Xu et al., Phys.Rev.Lett. {\bf 85}, 716 (2000).
\bibitem{dit01}M. Di Toro et al.,Nucl. Phys. {\bf A681}, 426c (2001);
V. Baran, M. Colonna, V. Greco, M. Di Toro, M.Zielinska Pfab\'e and
 H.H. Wolter, Nucl. Phys. {\bf A} (2002) in press.


\end{references}
\end{document}